\documentclass[prl,twocolumn,superscriptaddress]{revtex4}
\usepackage{float}
\usepackage{makeidx}
\usepackage{graphicx,amsmath}
\usepackage{CJK}

\begin{document}

\title{Spin-Orbit Coupled Degenerate Fermi Gases}

\author{Pengjun Wang\footnotemark}

\affiliation{State Key Laboratory of Quantum Optics and Quantum
Optics Devices, Institute of Opto-Electronics, Shanxi University,
Taiyuan 030006, P.R.China }

\author{Zeng-Qiang Yu\footnotemark}

\affiliation{Institute for Advanced Study, Tsinghua University,
Beijing, 100084,  P.R.China}

\author{Zhengkun Fu}

\affiliation{State Key Laboratory of Quantum Optics and Quantum
Optics Devices, Institute of Opto-Electronics, Shanxi University,
Taiyuan 030006,  P.R.China }

\author{Jiao Miao}

\affiliation{Institute for Advanced Study, Tsinghua University,
Beijing, 100084,  P.R.China}

\author{Lianghui Huang}

\affiliation{State Key Laboratory of Quantum Optics and Quantum
Optics Devices, Institute of Opto-Electronics, Shanxi University,
Taiyuan 030006,  P.R.China }

\author{Shijie Chai}

\affiliation{State Key Laboratory of Quantum Optics and Quantum
Optics Devices, Institute of Opto-Electronics, Shanxi University,
Taiyuan 030006,  P.R.China }

\author{ Hui Zhai}
\email{hzhai@mail.tsinghua.edu.cn} \affiliation{Institute for
Advanced Study, Tsinghua University, Beijing, 100084,  P.R.China}

\author{Jing Zhang}
\email{jzhang74@sxu.edu.cn; jzhang74@yahoo.com} \affiliation{State
Key Laboratory of Quantum Optics and Quantum Optics Devices,
Institute of Opto-Electronics, Shanxi University, Taiyuan 030006,
P.R.China }


\begin{abstract}

Spin-orbit coupling plays an increasingly important role in the
modern condensed matter physics. For instance, it gives birth to
topological insulators and topological superconductors. Quantum
simulation of spin-orbit coupling using ultracold Fermi gases will
offer opportunities to study these new phenomena in a more
controllable setting. Here we report the first experimental study of
a  spin-orbit coupled Fermi gas. We observe spin dephasing in spin
dynamics and momentum distribution asymmetry in the equilibrium
state as hallmarks of spin-orbit coupling. We also observe evidences
of Lifshitz transition where the topology of Fermi surfaces change.
This serves as an important first step toward finding Majorana
fermions in this system.

\end{abstract}

\maketitle

In the past decade, quantum simulations with ultracold
atoms have investigated many fascinating quantum phenomena in a
highly controllable and tunable way. Studies of ultracold Fermi
gases with resonant interaction have shed lights on understanding
strongly interacting fermion systems in nature, including
neutron stars  and quark-gluon plasma; simulating
Fermi Hubbard model with optical lattices helps to understand the
physical mechanism of unconventional high-Tc superconductors. However, until very recently, an important interaction has not been explored in cold atom systems, that is,
the spin-orbit (SO) coupling. Recently using two-photon Raman
process, SO coupled Bose-Einstein condensate has been realized in
the laboratory \cite{NIST}, which gives rise to new quantum phases
such as stripe superfluid \cite{Zhai,Ho}. In real materials, SO
coupling plays an important role in many physical systems over a
wide range of energy scale, from determining the nuclear structure
inside a nuclei, and the electronic structure inside an atom, to
giving birth to topological insulators in solid state materials
\cite{TI1,TI2}. Since all these systems are fermionic, from the
viewpoint of quantum simulation it is desirable to experimentally
realize SO coupled degenerate Fermi gases. The physical effects of
SO coupling in a degenerate Fermi gas are quite different from those
in a Bose system. In this work we report the first experimental
study of a SO coupled degenerate Fermi gas. Evidences of SO
coupling have been obtained from both the Raman induced quantum spin
dynamics and the spin-resolved momentum distribution. With SO
coupling, we also observe evidences for Lifshitz transition where the Fermi surface changes its topology as the density of fermion increases.
This progress will enable us to study stronger pairing and higher $T_c$
enhanced by SO coupling in resonant interacting Fermi gases
\cite{Vijay,Vijay_2,Yu,Hu} and topological insulator and topological
superfluid in a more flexible setup \cite{Zhang,Ian} in the near future.

In ultracold atom systems, SO coupling is generated through atom-light interaction induced artificial gauge fields,
and therefore it has two
advantages. First, both the strength and the configuration of SO
coupling are tunable via controlling the atom-light coupling;
Secondly, the coefficient of SO coupling is naturally on the same order of the inverse of laser wavelength, and in a gaseous system
it gives rise to a coupling strength as strong as the Fermi energy. In this regime SO coupling
dramatically changes the density-of-state at Fermi energy and the
topology of Fermi surface, which gives rise to many intriguing
phenomena in many-body systems
\cite{Vijay,Vijay_2,Yu,Hu,Zhang,Ian,review}, while such a regime is not easy
to access in conventional solid state materials.

\begin{figure}
\centerline{
\includegraphics[width=3in,height=4in]{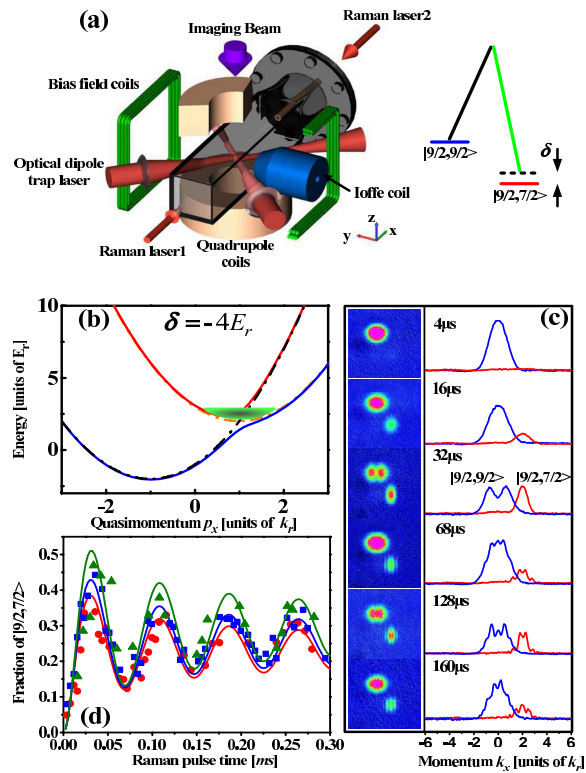}
} \vspace{0.1in} \caption{Experimental setup and Raman-induced
quantum spin dynamics: (a) Schematic of experimental setup and the
Raman coupling of two hyperfine levels of $^{40}$K. (b) The energy
dispersion with $\delta=-4E_{\text{r}}$. The system is initially
prepared with all atoms in $|9/2,9/2\rangle$ state. (c) The
population in $|9/2,7/2\rangle$ as a function of duration time of
Raman pulse. $k_{\text{F}}=1.9 k_{\text{r}}$ and
$T/T_{\text{F}}=0.30$ for red circles, $k_{\text{F}}=1.35
k_{\text{r}}$ and $T/T_{\text{F}}=0.35$ for blue squares,
$k_{\text{F}}=1.1 k_{\text{r}}$ and $T/T_{\text{F}}=0.29$ for green
triangles. The solid lines are theory curves with
$\Omega=1.52E_{\text{r}}$. (d) Time-of-flight image (left) and
integrated time-of-flight image (integrated along $\hat{y}$) at
different duration time for $|\uparrow\rangle$ (blue) and
$|\downarrow\rangle$ (red). The parameters are the same as blue ones
in (c) \label{Rabi} }
\end{figure}

In our experiment, a degenerate Fermi gas of $2\times 10^6$ $^{40}$K
atoms in the lowest hyperfine state $|F=9/2,m_{F}=9/2\rangle$ state is first prepared in an
optical dipole trap. The optical dipole trap is composed of two
horizontal crossed beams of $1064$ nm at $90^{o}$ along the $\hat{x}\pm
\hat{y}$ direction overlapped at the focus, as shown in Fig. \ref{Rabi}(a).
The temperature of the Fermi gas is about $0.3$-$0.4$ $T_{\rm F}$
($T_{\rm F}$ the Fermi temperature) when the trap frequency reaches
$2\pi\times(116, 116, 164)$ Hz along $(\hat{x},\hat{y},\hat{z})$
direction (see \textit{Methods} for details). We can change the trap
frequency adiabatically to achieve different fermion density. A pair
of Helmholtz coils (green ones in Fig. 1(a)) provides a homogeneous
bias magnetic field along $\hat{y}$ (quantization axis), which is
precisely controlled by a carefully designed scheme described in
Ref. \cite{ZhangJing} to reduce the magnetic field drift and the
magnetic noise.

The method we used to generate SO coupling is the same as reported
by the NIST group for the $^{87}$Rb Bose condensate \cite{NIST}. In
the $^{40}$K system, two spin-$1/2$ states are chosen as two
magnetic sublevels
$\mid\uparrow\rangle=|F=9/2,m_{\text{F}}=9/2\rangle$ and
$\mid\downarrow\rangle=|F=9/2,m_{\text{F}}=7/2\rangle$ of the
$F=9/2$ hyperfine level. They are coupled by a pair of Raman beams
with coupling strength $\Omega$. Two Raman lasers with the
wavelength $\lambda=773$ nm and the frequency difference $\omega$
counter-propagate along $\hat{x}$ axis and are linearly polarized
along $\hat{y}$ and $\hat{z}$ directions, respectively,
corresponding to $\pi$ and $\sigma$ polarization relative to
quantization axis $\hat{y}$ (as shown in Fig. 1(a)). The recoil
momentum $k_{\text{r}}=k_{0}\sin(\theta/2)$, and recoil energy
$E_{\text{r}}=k_{\text{r}}^{2}/2m = h\times 8.34$ kHz are taken as
natural momentum and energy units, where  $k_{0}=2\hbar\pi /\lambda$
and $\theta=180^{o}$ is the angle between two Raman beams. A Zeeman
shift $\omega_{\rm Z}/2\pi=10.4$ MHz between these two magnetic
sublevels is produced by the homogeneous bias magnetic field at $31$
G.  When the Raman coupling is at resonance (at $\omega/2\pi=10.4$
MHz and two-photon Raman detuning $\delta=\hbar(\omega_{\rm
Z}-\omega)\approx 0$), the detuning between
$|F=9/2,m_{\text{F}}=7/2\rangle$ and other magnetic sublevels like
$|F=9/2,m_{\text{F}}=5/2\rangle$ is about $h\times170$ kHz, which is
one order of magnitude larger than the Fermi energy. Thus we can
safely disregard other levels and treat this system as a spin-$1/2$
system. Same as in the boson experiment, this scheme generates an
effective single particle Hamiltonian as \cite{NIST}
\begin{equation}
  \mathcal{H}= \begin{pmatrix}{1\over 2m}( {\bf p}- k_{\rm r}\hat{\bf  e}_x)^2-{\delta\over 2} && {\Omega\over 2} \\ {\Omega\over 2} && {1\over 2m}( {\bf p}+ k_{\rm r}{\bf \hat e}_x)^2+{\delta\over 2} \end{pmatrix}\label{H0}
\end{equation}
Here, $ {\bf p}$ denotes the quasi-momentum of atoms, which relates to
the real momentum ${\bf k}$ as ${\bf k}= {\bf p}\mp k_{\rm r}\hat
{\bf e}_{ x}$ with $\mp$ for spin-up and down, respectively. This
Hamiltonian can be interpreted as an equal weight combination of
Rashba-type and Dresselhaus-type SO coupling \cite{NIST}. Finally,
before time-of-flight measurement, the Raman beams, optical dipole trap
and the homogeneous bias magnetic field are turned off abruptly at
the same time, and a magnetic field gradient along
$\hat{y}$ direction provided by Ioffe coil is turned on. Two
spin states are separated along $\hat{y}$ direction during the
time-of-flight due to the Stern-Gerlach effect, and imaging of atoms
along $\hat{z}$ direction after $12$ ms expansion
gives the momentum distribution for each spin component.

The main part of this paper is to study manifestation SO coupling in a Fermi gas system. We first study the Rabi oscillation between the two spin states
induced by the Raman coupling. This experiment, on one hand,
measures the coupling strength $\Omega$; and on the other hand,
clearly manifests different effect of SO coupling compared to boson
system. All atoms are initially prepared in the
$\mid\uparrow\rangle$ state. The homogeneous bias magnetic field is
ramped to a certain value so that $\delta=-4E_{\text{r}}$, that is,
the ${\bf k}=0$ state of $\mid\uparrow\rangle$ component is at
resonance with ${\bf k}=2k_{\text{r}}\hat {\bf e}_{ x}$ state of
$\mid\downarrow\rangle$ component, as shown in Fig. \ref{Rabi}(b).
Then we apply a Raman pulse to the system, and measure the spin population for different duration time of the
Raman pulse. Similar experiment in boson system yields an undamped
and completely periodic oscillation, which can be well described by
a sinusoidal function with frequency $\Omega$ \cite{IanPRL}. This is
because for bosons, macroscopic number of atoms occupy the resonant
${\bf k}=0$ mode, and therefore there is a single Rabi frequency
determined by the Raman coupling only. While for fermions, due to
the Pauli exclusion principle, atoms occupy different momentum
states. Precisely due to the effect of SO coupling, the coupling
between the two spin states and the resulting energy splitting are
momentum dependent, and atoms in different momentum states oscillate
with different frequencies (as shown explicitly in Eq.
(\ref{Rabitheory}) later). Hence, dephasing naturally occurs and the
oscillation will be inevitably damped after several oscillation
periods. This process is very similar to the spin dynamics of a spin
polarized current ejected into a semiconductor. For semiconductor spintronics, one would like to have a polarized spin current,
however, because different electron has different velocity and thus their spins precess in different way, the current will be unpolarized.
Such spin dynamics has been extensively studied during recent decades \cite{Wu}. In our case, we can observe momentum-resolved
spin dynamics with Stern-Gerlach technique and therefore we can clearly reveal this physical process. In the
Fig. \ref{Rabi}(c), we show the momentum distribution for both spin
components at several different moments. One can see the
multiple-peaks feature in momentum distribution in Fig. \ref{Rabi}(d), which clearly shows the out-of-phase oscillation for different momentum
states.

Unlike bosons, the Raman coupling $\Omega$ cannot be read out
directly from the period of oscillation. In fact, the frequency
depends on both atoms density and temperature, and it is generally
smaller than $\Omega$ due to the effect of averaging over different
momentum states. To determine the value of $\Omega$ from the
measurements, we fix Raman coupling and vary atoms density by
changing the total number of fermions or the trapping frequency, and
we obtain several different oscillation curve, as shown in Fig.
\ref{Rabi}(d). Then we fit them to the theory with a single fitting
parameter $\Omega$. Theoretically, for a non-interacting system, the
population of $\mid\downarrow\rangle$ component is given by
\begin{eqnarray}
n_{\downarrow}({\bf k}+2 k_{\text{r}}\hat{{\bf e}}_x, {\bf r},
t)=n_{\uparrow}({\bf k}, {\bf r}, 0)\frac{1}{1+\left(\frac{2 k_x
k_\text{r}}{\Omega m}\right)^2}\times \nonumber \\
\sin^2\sqrt{(k_x k_\text{r}/m)^2+\Omega^2/4}\,t, \label{Rabitheory}
\end{eqnarray}
where $t$ is the duration time of Raman pulse, $n_{\uparrow}({\bf
k},{\bf r},0)$ is the equilibrium distribution of the initial state
in local density approximation, and temperature of initial cloud is
determined by fitting the time-of-flight image to momentum
distribution of free fermions in a harmonic trap. From Eq.
(\ref{Rabitheory}) one can see that the momentum distribution along
$\hat{x}$ direction of $\mid\downarrow\rangle$ component is always
symmetric respect to $2k_{\text{r}}$ at any time, and the
experimental data is indeed the case, as shown in Fig.
\ref{Rabi}(c). The theoretical expectation of the total population
in $\mid\downarrow\rangle$ component is given by
$N_{\downarrow}(t)=\int {d^3{\bf k}d^3{\bf r}} n_{\downarrow}({\bf
k},{\bf r},t)$, and in Fig. \ref{Rabi}(d), one can see there is an
excellent agreement between the experiment data and theory, from
which we determine $\Omega=1.52(5) E_{\rm r}$. Since our current
experiment is performed in the weakly interacting regime with
$s$-wave scattering length $a_{\text{s}}=169 a_0$, we have verified
that the interaction effect is negligible (see {\it Method} for
details).

\begin{figure}
\centerline{
\includegraphics[width=3in]{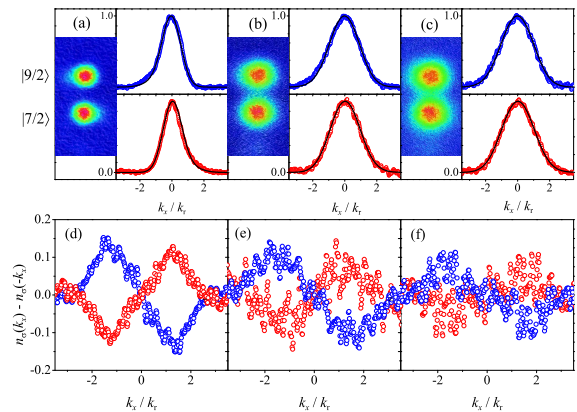}
} \vspace{0.1in} \caption{Momentum distribution asymmetry as a
hallmark of SO coupling: (a-c) time-of-flight measurement of
momentum distribution for both $|\uparrow\rangle$ (blue) and
$|\downarrow\rangle$ (red). Solid lines are theory curves. (a)
$k_{\text{F}}=0.9 k_{\text{r}}$ and $T/T_{\text{F}}=0.8$ (b)
$k_{\text{F}}=1.6 k_{\text{r}}$ and $T/T_{\text{F}}=0.63$; (c)
$k_{\text{F}}=1.8 k_{\text{r}}$ and $T/T_{\text{F}}=0.57$. (d-f):
plot of integrated momentum distribution $n_{\sigma}({\bf
k})-n_{\sigma}({\bf -k})$ for the case of (a-c).
\label{distribution} }
\end{figure}

Next, we focus on the case with $\delta=0$, and study the
momentum distribution in the equilibrium state. We first transfer
half of $^{40}$K atoms from $\mid\downarrow\rangle$ to
$\mid\uparrow\rangle$ using radio frequency sweep within $100$ ms.
Then the Raman coupling strength is ramped up adiabatically in $100$
ms from zero to its final value and the system is held for another
$50$ ms before time-of-flight measurement. We have also varied the
holding time and find the momentum distribution does not change, and
thus we conclude that the system has reached equilibrium in the
presence of SO coupling. Since SO coupling breaks spatial
reflectional symmetry ($x\rightarrow -x$ and $k_x\rightarrow -k_x$),
the momentum distribution for each spin component will be
asymmetric, i.e. $n_\sigma({\bf k})\neq n_\sigma({\bf -k})$, with
$\sigma=\uparrow,\downarrow$; While on the other hand, when
$\delta=0$ the system still preserves time-reversal symmetry, which
requires $n_{\uparrow}({\bf k})=n_{\downarrow}(-{\bf k})$. The
asymmetry can be clearly seen in the spin-resolved time-of-flight
images and integrated distributions displayed in Fig.
\ref{distribution}(a) and (b), where the fermion density is relatively low. While it becomes less significant when the fermion density becomes higher, as shown in Fig. \ref{distribution}(c),
because the strength of SO coupling is relatively weaker compared to the Fermi energy.
In Fig. \ref{distribution}(a-c) the integrated momentum distribution is
fitted by the theoretical calculation to determine the temperature and
the chemical potential at the center of the trap. (See \textit
{Method} for detail). We find that the Raman lasers indeed cause
additional heating to the cloud. Nevertheless, the temperature we
find is within the range of $0.5-0.8 T_{\text{F}}$ \cite{notation},
which is still below degenerate temperature. In Fig.
\ref{distribution}(d-f), we also show
$n_{\sigma}(k_x)-n_{\sigma}(-k_x)$ to reveal the distribution
asymmetry more clearly. These are smoking gun evidences of a degenerate Fermi gas with SO coupling.

\begin{figure}
\centerline{
\includegraphics[width=3in,height=4in]{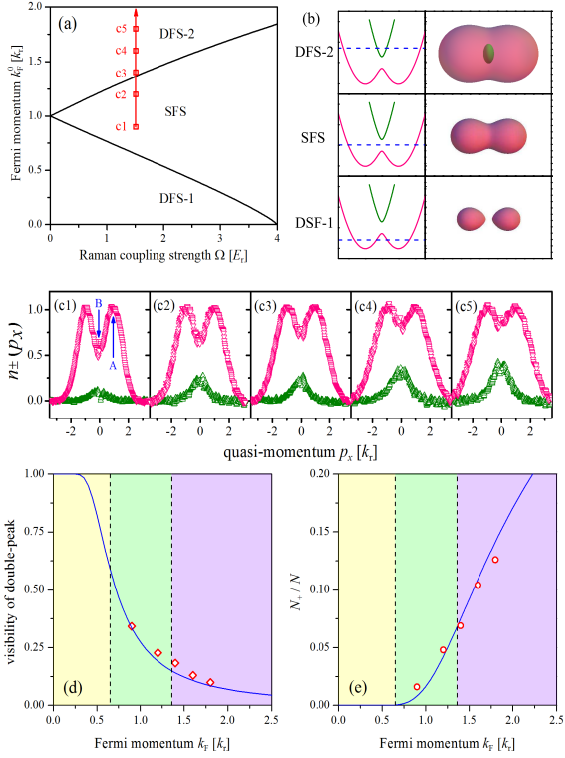}
} \vspace{0.1in} \caption{Topological change of Fermi surface and
Lifshitz transition: (a) Theoretical phase diagram at $T=0$.
$k^0_{\text{F}}=\hbar(3\pi^2 n)^{1/3}$. ``SFS" means single Fermi
surface. ``DFS" means double Fermi surface. (b) Illustration of
different topology of Fermi surfaces as Fermi energy increases. The
single particle energy dispersion is drawn for small $\Omega$, in
which red and green lines are for $s=-1$ and $s=1$ helicity
branches, respectively. Dashed blue line is the chemical potential.
Red and green surfaces are Fermi surfaces for $s=-1$ helicity branch
and $s=1$ helicity branch, respectively. (c) Quasi-momentum
distribution in the helicity bases. Red and green points are
distributions for $s=-1$ and $s=1$ helicity branches, respectively.
$k_{\text{F}}=0.9 k_{\text{r}}$, $T/T_{\text{F}}=0.80$ for (c1);
$k_{\text{F}}=1.2 k_{\text{r}}$, $T/T_{\text{F}}=0.69$ for (c2);
$k_{\text{F}}=1.4 k_{\text{r}}$, $T/T_{\text{F}}=0.61$ for (c3);
$k_{\text{F}}=1.6 k_{\text{r}}$, $T/T_{\text{F}}=0.63$ for (c4);
$k_{\text{F}}=1.8 k_{\text{r}}$, $T/T_{\text{F}}=0.57$ for (c5). All
these points are marked on phase diagram in (a). (d) Visibility
$v=(n_{\text{A}}-n_{\text{B}})/(n_{\text{A}}+n_{\text{B}})$
decreases as $k_{\text{F}}/k_{\text{r}}$ increases (A and B points
are marked in (c1)). (e) Atom number population in $s=1$ helicity
branch $N_{+}/N$ increases as $k_{\text{F}}/k_{\text{r}}$ increases
increases. In both (d) and (e), the blue solid line is a theoretical
curve with $T/T_{\text{F}}=0.65$, and the background color indicates
three different phases in the phase diagram. \label{transition} }
\end{figure}

With SO coupling, the single particle spectra of Eq. \ref{H0} are dramatically
changed from two parabolic dispersions into two helicity branches as
shown in the inset of Fig. \ref{transition}(a). Here, two different
branches are eigenstates of ``helicity" $\hat{s}$ and the
``helicity" operator describes whether spin
$\boldsymbol{\sigma}_{{\bf p}}$ is parallel or anti-parallel to the
``effective Zeeman field"  ${\bf h}_{{\bf p}}=(-\Omega, 0,
k_{\text{r}}p_x/m+\delta)$ at each momentum, i.e.
$\hat{s}=\boldsymbol{\sigma}_{{\bf p}}\cdot{\bf h}_{{\bf
p}}/|\boldsymbol{\sigma}_{{\bf p}}\cdot{\bf h}_{{\bf p}}|$. $s=1$
for the upper branch and $s=-1$ for the lower branch. The topology
of Fermi surface exhibits two transitions as the atoms density
varies. At sufficient low density, it contains two disjointed Fermi
surfaces with $s=-1$, and they gradually merge into a single Fermi surface as
the density increases to $n_{\text{c}1}$. Finally a new small Fermi
surface appears at the center of large Fermi surface when density
further increases and fermions begin to occupy $s=1$ helicity branch
at $n_{\text{c}2}$. A theoretical ground state phase diagram for the
uniform system is shown in Fig. \ref{transition}(a), and an
illustration of the Fermi surfaces at different density are shown in
Fig. \ref{transition}(b). Across the phase boundaries, the system
experiences Lifshitz transitions as density increases \cite{Lifshitz}, which is a unique property in a Fermi gas due to Pauli principle. At sufficiently low temperature,
the derivative of the thermodynamic quantities like the
compressibility will exhibit singularity in the critical regime
around the transition point.

Rigorously speaking Lifshitz transition only exists at zero temperature and at finite temperature it becomes a crossover. However, we can still obtain several consistent features that supports the existence of such a transition at zero temperature. We fix the Raman coupling and
vary the atoms density at the center of the trap by controlling
total fermion number or trap frequency, as indicated by the red
arrow in Fig. \ref{transition}(a). The quasi-momentum distribution
in the helicity bases can be obtained from a transformation of
momentum distribution in spin bases as follows (See {\textit
{Method}} for the definition of $u_{\bf  p}$ and $v_{\bf  p}$):
\begin{align}
  n_{+}({\bf  p}) & ={u_{\bf  p}^2 n_{\uparrow}({\bf  p}-\hbar k_{\rm r}{\bf \hat e}_x)- v_{\bf  p}^2 n_{\downarrow}({\bf  p} + \hbar k_{\rm r}{\bf \hat e}_x) \over u_{\bf  p}^2 - v_{\bf  p}^2} \\
  n_{-}({\bf  p}) & ={v_{\bf  p}^2 n_{\uparrow}({\bf  p}- \hbar k_{\rm r}{\bf \hat e}_x)- u_{\bf  p}^2 n_{\downarrow}({\bf  p} + \hbar k_{\rm r}{\bf \hat e}_x) \over v_{\bf  p}^2 - u_{\bf  p}^2}
\end{align}
In Fig. \ref{transition}(c1-c5), we plot the quasi-momentum
distribution in the helicity bases for different atoms density. At
the lowest density, the $s=1$ helicity branch is nearly unoccupied,
which is consistent with that the Fermi surface is below $s=1$
helicity branch. The quasi-momentum distribution of the $s=-1$
helicity branch exhibits clearly a double-peak structure, which
reveals the emergence of two disjointed Fermi surfaces at $s=-1$
helicity branch. As density increases, the double-peak feature
gradually disappears, indicating the Fermi surface of $s=-1$
helicity branch finally becomes a single elongated one, as the top
one in Fig. \ref{transition}(b). Here we define a quality of
visibility
$v=(n_{\text{A}}-n_{\text{B}})/(n_{\text{A}}+n_{\text{B}})$, where
$n_{\text{A}}$ is the density of $s=-1$ branch at the peak and
$n_{\text{B}}$ is the density at the dip between two peaks.
Theoretically one expects $v$ approaches unity at low density regime
and approaches zero at high density regime. In Fig
\ref{transition}(d) we show that our data decreases as density
increases and agrees very well with a theoretical curve with fixed
temperature $T/T_{\text{F}}=0.65$. Moreover, across the phase
boundary between SFS and DFS-1, there will be a significant increase
of population on $s=1$ helicity branch.  In Fig. \ref{transition}
(e), the fraction of atom number population at $s=1$ helicity branch
is plotted as a function of Fermi momentum $k_{\rm F}$, which grows
up rapidly nearby the critical point predicted in zero-temperature
phase diagram. The blue solid line is a theoretical calculation for
$N_{+}/N$ with $T/T_{\text{F}}=0.65$, and the small deviation
between the data and this line is due to the temperature variation
between different measurements. Both two features are consistent
with a Lifishitz transition. Recently, topological change of Fermi
surface and Lifshitz transition have also been studied in Fermi gas
in honeycomb optical lattices, where single particle spectrum
exhibits Dirac point behavior \cite{Dirac}. A more accurate
experimental determination of Lifshitz transition point requires
more well control of temperature in presence of Raman lasers and the
information of local equation-of-state. In the near future, we also
plan to bring the system close to a Feshbach resonance where the
$s$-wave interaction becomes strongly attractive, and we will further
cool the system below the superfluid transition temperature. There
we expect to find Majorana fermion modes at the phase boundaries
when Fermi surface topology changes
\cite{Majorana,Majorana1,Majorana2}.

In summary, we have for the first time studied effects of spin-orbit coupling in a ultracold atomic Fermi gas. We study Raman induced spin
dynamics and reveal clearly the physical process of spin dephasing
due to SO coupling. We measure the momentum distribution in the
equilibrium state and find the momentum distribution asymmetry due
to the broken of the reflection symmetry caused by SO coupling. We also find
evidences for the change of Fermi surface topology and Lifshitz
transition from the quasi-momentum distribution in the helicity
bases. All these features are unique manifestations of SO coupling
in fermionic system, which are absent in previously realized SO
coupled boson condensate \cite{NIST}. This research opens up an
avenue toward rich physics of SO coupled Fermi gases.

{\bf Methods.}

{\it Experimental setup.} The experimental setup (see Fig. 1) is the
same as discussed in \cite{twenty-four,twenty-five,twenty-five1},
which employs a mixture of $^{87}$Rb and $^{40}$K atoms. In short, a
mixture of $1\times10^{7}$ $^{87}$Rb atoms at the spin state
$|F=2,m_{F}=2\rangle$ and $4\times10^{6}$ $^{40}$K atoms at
$|F=9/2,m_{F}=9/2\rangle$ are precooled to 1.5 $\mu K$ by
radio-frequency evaporative cooling in a quadrupole-Ioffe
configuration (QUIC) trap, and then are transported to the center of
the glass cell \cite{twenty-six} in favor of optical access. Both
species are loaded into the optical dipole trap and further
evaporatively cooled in the optical dipole trap by evaporative
cooling. At the end of above process, the $2\times10^{6}$ $^{40}$K
atoms reaches quantum degeneracy at $\sim 0.3T_{\text{F}}$.  We use
a resonant laser beam for $0.03$ ms to remove the Rb atoms without
heating $^{40}$K atoms.

Two laser beams for generating Raman coupling are extracted from a
Ti-sapphire laser operating at the wavelength of $773$ nm with the
narrow linewidth single-frequency and focused at the position of the
atomic clouds with $1/e^{2}$ radii of 200 $\mu$m. Two Raman beams
are frequency-shifted with -100 MHz and -110.4 MHz by two
single-pass acousto-optic modulators (AOM) respectively, and then
are coupled into two polarization maintaining single-mode fibers in
order to increase stability of the beam pointing and obtain better
beam-profile quality. The frequency difference $\omega$ of two
signal generators for two AOMs is phase-locked by a source locking
CW microwave frequency counters. To enhance the intensity stability
of the two Raman beams, a small fraction of the light is sent into a
photodiode and the regulator is used for comparing the intensity
measured by the photodiode to a set voltage value from the computer.
The non-zero error signal is compensated by adjusting the
radio-frequency power in the AOM in front of the optical fiber.

{\it Calculating Momentum Distribution in a Trap.} The energy
eigenstates for the single particle Hamiltonian in Eq. (1) are the
dressed states usually denoted as helicity bases with $s=\pm 1$, with
$|+,{\bf p}\rangle= u_{\bf  p} |\uparrow, {\bf
p}\rangle +v_{\bf  p} |\downarrow, {\bf  p}\rangle $ and
$|-,{\bf p}\rangle= v_{\bf  p} |\uparrow, {\bf
p}\rangle -u_{\bf  p} |\downarrow, {\bf  p}\rangle $,
where the coefficients $u_{\bf  p}^2=1-v_{\bf  p}^2 =
{1\over 2}\big[1+{p_x v_{\rm r}-\delta/2\over \sqrt{(p_xv_{\rm
r}-\delta/2)^2+\Omega^2/4}}\big]$ are only functions of $ p_x$
with $v_{\rm r}= k_{\rm r}/m$. The energy dispersion for each
branch of these dressed states are given by $E_{\pm,{\bf
p}}=( {\bf p}^2+k_{\rm r}^2)/(2m)\pm \sqrt{( p_xv_{\rm
r}-\delta/2)^2+\Omega^2/4}$.

The trapping potential $V({\bf r})$ is taken into account by local density
approximation. Even in presence of Raman coupling, what can be measured in the time-of-flight image with Stern-Gerlach technique are still momentum distributions of original hyperfine spin states. Theoretically they are given by
\begin{align}
  n_{\uparrow}({\bf  p}- k_{\rm r}{\bf \hat e}_x)=\int{d^3{\bf r}\over (2\pi\hbar)^3}\Big[u_{\bf  p}^2 f(E_{{\bf  p},+};{\bf r}) + v_{\bf  p}^2 f(E_{{\bf  p},-};{\bf r})\Big] \nonumber \\
   n_{\downarrow}({\bf  p}+ k_{\rm r}{\bf \hat e}_x)=\int{d^3{\bf r}\over (2\pi\hbar)^3}\Big[v_{\bf  p}^2 f(E_{{\bf  p},+};{\bf r}) + u_{\bf p}^2 f(E_{{\bf  p},-};{\bf r})\Big] \nonumber
\end{align}
where $f(E;{\bf r})$ is the Fermi-Dirac distribution function with a
local chemical potential $\mu({\bf r})=\mu-V({\bf r})$, and $\mu$ is
determined by the equation for total particle number
\begin{align}
  N=\int d^3{\bf p}\Big[n_{\uparrow}({\bf p})+n_{\downarrow}({\bf p})\Big]
\end{align}
By fitting the measured data  to the theoretical curve of integrated
momentum profile $n_{\rm 1d,\sigma}(p_x)=\int{\rm d}p_ydp_z\, n_{\sigma}({\bf
p})$ with $\sigma=\uparrow$ or $\downarrow$, we obtain the temperature of atoms in the experiment.

{\it Interaction Induced Change of Rabi Frequency.}
We shall also compare this experiment to the clock experiment in
Fermi gases \cite{Ye}, where the coupling between two components is
momentum independent if the light intensity is assumed to be
uniform. In that case, all atoms undergo Rabi oscillation with exact
same frequency, and therefore remain as identical particles during
the process. Hence, the $s$-wave interaction plays no role except
when the spatial inhomogeneity is taken into account \cite{Ye}. In
contrast, in our case different atoms oscillate with different
frequencies, thus they immediately become distinguishable particles
as the oscillation starts, and can interact with each other via
$s$-wave collisions. Since our experiment is performed in the weakly interacting regime, we
include the interaction effect with mean-field theory.

For a non-interacting system, the equation of motion for a spin ${\bf S}_{\bf p}$ with quasi-momentum ${\bf  p}$  is given by
\begin{align}
  {\partial \over \partial t}{\bf S}_{\bf  p} = {\bf S}_{\bf  p}\times {\bf h}_{\bf  p}
\end{align}
where ${\bf h}_{\bf  p}=(-\Omega, 0,k_{\rm r} p_x/m+\delta )$ is the
momentum dependent ``effective magnetic field". The frequency of the
spin rotation is given by $|{\bf h}_{\bf  p}|=\sqrt{( p_xk_{\rm
r}/m+\delta)^2+\Omega^2}$, and for $\delta=-4E_{\rm r}$, it is just
the frequency of population oscillation as shown in Eq.
(\ref{Rabitheory}). For a weakly-interacting system, the interaction
can be approximated by $H_{\rm int}={g\over V}{\bf S}\cdot {\bf S}$
at mean-field level, where $g=4\pi\hbar^2 a_s/m$ is the coupling
constant, $a_s$ is the $s$-wave scattering length, and ${\bf
S}=\sum_{\bf p}{\bf S}_{\bf p}$ is the total spin. Thus, an
additional mean-field term appears in the equation of motion,
\begin{align}
  {\partial \over \partial t}{\bf S}_{\bf  p} = {\bf S}_{\bf  p}\times \big({\bf h}_{\bf  p}+2{g\over V}{\bf S}\big).
\end{align}
where the solution of ${\bf S}_{\bf p}$ must be determined
self-consistently. We numerically solve this equation of motion to
determine the spin dynamics. Using the experimental parameters, we
find the frequency shift is only a few percent of $\Omega$, which is
beyond the measurement resolution of current experiment.

{\bf Acknowledgements.} We would like to thank Zhenhua Yu, Cheng
Chin, Tin-Lun Ho and Sandy Fetter for helpful discussions. This
research is supported by National Basic Research Program of China
(Grant No. 2011CB921601, 2010CB923103, 2011CB921500), NSFC (Grant
No. 10725416, 61121064, 11004118, 11174176), DPFMEC (Approval No.
20111401130001) and Tsinghua University Initiative Scientific
Research Program.



\begin{thebibliography}{99}
\bibitem{notation}
Here, we define Fermi energy and Fermi temperature in the
conventional way as the trapped Fermi gas without SO coupling,
$E_{\rm F}=k_{\rm B}T_{\rm F}= k_{\rm F}^2 /(2m)$, with $k_{\rm
F}=(24N)^{1/6}\sqrt{\hbar m\omega_{\rm ho}}$ the Fermi momentum and
$\omega_{\rm ho}$ the average trap frequency.
\end{thebibliography}
\end{document}